\documentclass[iop]{emulateapj}
 \usepackage{amsmath, amsthm, amssymb,amsfonts,color}
\usepackage[english]{babel}

\begin{document}

\graphicspath{{graphs/}}

\title{The parametric instability of  Alfv\'en waves: effects of temperature anisotropy }

\author{Anna Tenerani, Marco Velli }
\affiliation{Department of Earth, Planetary, and Space Sciences, University of California, Los Angeles, CA 90095}
\email{annatenerani@epss.ucla.edu, mvelli@ucla.edu}
%\author{Marco Velli}
%\affiliation{Department of Earth, Planetary, and Space Sciences, University of California, Los Angeles, CA}
\author{Petr Hellinger}
\affiliation{Astronomical Institute, CAS, Bocni II/1401, CZ-14100 Prague, Czech Republic}

\begin{abstract}
We study the stability  of large amplitude, circularly polarized Alfv\'en waves in an anisotropic plasma  described by the double-adiabatic/CGL closure,  and in particular the effect of a background thermal pressure anisotropy on the well-known properties of Alfv\'en wave parametric decay in Magnetohydrodynamics (MHD). Anisotropy  allows instability over  a much wider range of values of parallel plasma beta ($\beta_\parallel$) when  $\xi=p_{0\bot}/p_{0\parallel} > 1$. When the pressure anisotropy exceeds a critical value, $\xi\geq\xi^*$ with $\xi^*\simeq2.7$, there is a new regime in which  the parametric instability is no longer quenched at high $\beta_\parallel$ and in the limit $\beta_\parallel\gg1$ the growth rate becomes independent of  $\beta_\parallel$. In the opposite case of $\xi< \xi^*$, the instability is strongly suppressed with increasing parallel plasma beta, similarly to  the MHD case. We analyze marginal stability conditions for parametric decay in the $(\xi,\,\beta_\parallel)$ parameter space, and discuss possible implications  for Alfv\'enic turbulence in the solar wind. 
\end{abstract}


\maketitle
%\begin{article}

\section{Introduction}
Alfv\'en waves are ubiquitous in magnetized, astrophysical plasmas. They are invoked to explain the heating of stellar coronae and the acceleration of stellar winds, as well as the origin and formation of galactic and extragalactic jets~\citep{pereira}. Since the first in-situ measurements~\citep{unti}, Alfv\'enic fluctuations have been commonly observed in the solar wind, especially in the fast streams possibly originating from coronal holes,  from a few AU  all the way down to 0.3~AU. 

In the solar wind, Alfv\'enic fluctuations dominate the low frequency part of the fluctuation energy spectrum, with  frequency $f$ in the range $f\simeq10^{-4}-10^{-2}$~Hz. They appear to be mainly propagating outwards from the Sun and yet display a power-law spectrum that evolves with heliocentric distance~\citep{carbone}. One of the most remarkable  properties observed during these so-called Alfv\'enic periods is that the amplitude of the magnetic field fluctuations $\delta {\bf B}$  is of the same order of the average, larger scale magnetic field ${\bf B}_0$, i.e. they satisfy  $|\delta{\bf B}|/|{\bf B}_0|\sim 1$. Despite such large excursions,  the magnitude of the total magnetic field $|{\bf B}|$ remains relatively constant with negligible associated compressive fluctuations~\citep{matteini_2014, matteini_2015}. How is this Alfv\'enic turbulent state  achieved in the solar wind, and what is its  dynamical role in  solar wind heating and acceleration remain fundamental open questions of Heliophysics and Astrophysics in general. 

Large amplitude Alfv\'enic fluctuations with constant total (magnetic+kinetic) pressure constitute an exact nonlinear state in Magnetohydrodynamics (MHD). However, both theory~\citep{galeev_sov_phys_1963, derby} and numerical simulations~\citep{DelZanna_AA_2001} show that such state is unstable and that  Alfv\'enic fluctuations evolve by coupling with  compressive and Alfv\'en modes. This instability, called parametric instability or parametric decay, has proven to be robust and not significantly affected by wave polarization or propagation direction~\citep{DelZanna_GRL_2001, matteini_GRL_2010}. Even in the presence of a broad spectrum of frequencies, where one might expect a quenching due to the difficulty of maintaining resonance conditions, parametric decay has been shown to survive~\citep{malara_Phys_fluids_1996,malara2}. The parametric instability is more efficient at low values of the plasma beta (thermal to magnetic pressure ratio), $\beta\ll1$. In this case, a forward propagating Alfv\'en wave with wave number $k_0$, the  mother wave, decays in two daughter waves: a forward sound wave with wave number $k_s\simeq 3/2k_0$ and a low frequency backward Alfv\'en wave with wave number $k_a=k_0-k_s\simeq -k_0/2$. Such a decay tends to a four-wave interaction at larger $\beta$, including  coupling with a forward  Alfv\'en wave. However,  parametric decay is strongly stabilized at values of  $\beta$ of order unity and beyond, unstable waves having amplitudes that scale as a positive power of $\beta$. 

The solar wind expansion has been  taken into account to investigate how it affects the parametric instability of waves launched from regions close to the Sun~\citep{anna1}. It was shown that the  solar wind expansion stabilizes the frequencies  with  growth rate of the order or smaller than the expansion rate (frequencies of about $10^{-4}$~Hz), whereas the higher frequencies are unstable and decay during their propagation in the outer corona and inner heliosphere. In the solar wind, however,  density fluctuations are extremely weak during Alfv\'enic periods, and the absence of signatures of Alfv\'en wave decay is mysterious.  Dispersive effects due to finite ion inertial length or  ion Larmor radius introduce a richer dynamics by breaking the symmetry between right and left hand polarizations and by allowing also modulational instabilities~\citep{sakai, wong}. The latter usually arise from the coupling of the mother wave with two forward daughter Alfv\'en waves, one with frequency and wavenumber greater than that of the mother wave,  and one with frequency and wavenumber smaller than that of the mother wave. In general, dispersion allows both modulational and parametric decay in regions where $\beta>1$, extending the range of unstable modes~\citep{hamabata_93a,vasquez_1995,hollweg,araneda_1998, nariyuki,araneda_2008}.  

So far, research on the stability of large amplitude  Alfv\'en waves and their nonlinear evolution has focused on (isotropic) plasmas in thermodynamic equilibrium. On the other hand, the weakly collisional solar wind displays several non-thermal features in the particle (proton) velocity distribution functions (PDF), especially in the fast streams~\citep{marsch}.  In particular, PDFs are often  characterized by thermal anisotropies that vary with heliocentric distance most probably due to a combination of kinetic instabilities and  expansion effects, which has not been described in detail. The PDFs are typically observed to be stable or marginally stable with respect to velocity-space instabilities driven by temperature anisotropies~\citep{petr_2006,bale,matteini_JGR_2013}.  

Here we study the stability of large amplitude, circularly polarized Alfv\'en waves propagating in an anisotropic plasma by adopting the one fluid double adiabatic description of the plasma~\citep{cgl}. We analyze how the background thermal anisotropy affects the properties of the parametric decay instability and we discuss under which conditions large amplitude Alfv\'en waves may be stable or not in the solar wind by providing diagrams in parameter space defined by the anisotropy vs. parallel plasma beta. Although we neglect {dispersive, Landau damping and other kinetic effects}, the present analysis is a first necessary step to understand the evolution of long wavelength Alfv\'en waves in the presence of velocity space anisotropies. We defer to later work the study of dispersion and kinetic effects.  We note that a more limited study but in a similar vein was presented in \citet{hamabata}. However, because of some discrepancies in the relevant equations, we will not discuss that work here.

The remainder of the paper is organized as follows: in section~\ref{model} we define the initial configuration and address the general properties of the nonlinear wave equation resulting from the CGL model; in section~\ref{results} we study parametric decay of Alfv\'en waves in an anisotropic plasma and in section~\ref{disc} we discuss possible implications for Alfv\'en waves in the solar wind; in section~\ref{sum} we summarize our results.

\section{Model equations and background configuration}
\label{model}
In this study we neglect dispersive effects due to finite Larmor radius (FLR) and ion inertial length by considering long wavelength and low frequency waves, so that the one fluid double adiabatic description of the plasma is suitable~{\citep{abraham,ferriere,cerri, hunana,dds}}. In general, different assumptions  may be employed to close the hierarchy of equations obtained by taking the moments in velocity space of the Vlasov-Maxwell equations, each of them leading to different fluid models with approximated FLR and Landau damping effects~\citep{passot,sulem,dds_2016}. For the purpose of this paper we adopt the well-known CGL framework in which not only dispersion, but also heat flux effects are entirely neglected~\citep{cgl}. Despite these simplifying assumptions, the CGL model  provides a good description of some  effects due to thermal pressure anisotropy at the large scales, while a proper description of the dynamics at smaller scales would best require { more sophisticated fluid models that include FLR effects~\citep{hunana_2017} or even} a full kinetic treatment. Ion-acoustic Landau damping may be modeled phenomenologically by adding an appropriate {drag} term to the longitudinal component of the momentum equation. This approach was employed to study how Landau damping affects parametric and modulational instabilities in fluid models, and it was shown to lead to results consistent with the outcome of hybrid simulations~\citep{vasquez_1995,gomberoff}. In particular, Landau damping reduces the growth rate of parametric decay at most by a factor of order two. For this reason, we have decided not to include it here, although it may be of interest to inspect its effects in future works devoted to more general, non-monochromatic waves~\citep{cohen}.

The  CGL model is given by the following set of equations, 
\begin{equation}
\frac{\partial\rho}{\partial t}+\boldsymbol\nabla\cdot{\bf (\rho u)}= 0
\label{CGL_eq0}  
\end{equation}
\begin{equation}
\begin{split}
&\rho\left(\frac{\partial {\bf u}}{\partial t}+{\bf u}\cdot\boldsymbol{\nabla}{\bf u}\right)=-\boldsymbol{\nabla}\left(p_\bot+ \frac{B^2}{8\pi}\right)+\\
&\frac{1}{4\pi}{\bf B}\cdot \boldsymbol{\nabla}{\bf B}+\boldsymbol\nabla\cdot\left( {\bf \hat b\hat b}\Delta p \right),
\end{split}
\end{equation}
\begin{equation}
\frac{\partial {\bf B}}{\partial t}=\boldsymbol{\nabla}\times({\bf u\times B}),
\end{equation}
\begin{equation}
\left(\frac{\partial }{\partial t}+{\bf u}\cdot\boldsymbol{\nabla}\right)\left( \frac{p_\parallel B^2}{\rho^3} \right)=0,
\label{CGL_eq1}  
\end{equation}
\begin{equation}
\left(\frac{\partial }{\partial t}+{\bf u}\cdot\boldsymbol{\nabla}\right)\left( \frac{p_\bot }{\rho B} \right)=0,
\label{CGL_eq2}  
\end{equation}
where ${\bf \hat b}={\bf B}/B$ is the unit vector along the magnetic field, $B=|{\bf B}|$ is the magnetic field magnitude and $\Delta p=p_\bot-p_\parallel$, perpendicular and parallel referring to the  magnetic field ${\bf B}$. 

Just as in MHD, perfectly correlated finite amplitude transverse magnetic and velocity fluctuations are an exact nonlinear state, or dynamical equilibrium, of Eqs.~(\ref{CGL_eq0})--(\ref{CGL_eq2}), provided the total (kinetic+magnetic) pressure is homogeneous (constant in space). The (one-dimensional) wave equation resulting from the CGL model, Eq.~(\ref{nl2}) below, however contains a nonlinear term which is absent in  MHD, in that the fluctuation propagation speed depends on the magnetic pressure of the fluctuation itself. This nonlinearity has important effects: it regularizes firehose unstable fluctuations that would grow without bounds in the linear approximation, and allows a class of nonlinear states that cannot be reduced to  a simple superposition of monochromatic waves. Depending upon the initial conditions, this class of solutions includes as particular case circularly polarized  Alfv\'en waves (i.e. temporally constant amplitude waves). 

Although many initial conditions can in principle be considered to study parametric decay, in this work we focus  on the stability of circularly polarized Alfv\'en waves. We review this specific case below, followed by a discussion on the evolution of fluctuations in the firehose unstable regime and on the  properties of the nonlinear wave equation.     

\subsection{Nonlinear circularly polarized Alfv\'en waves in anisotropic plasmas}
We begin by seeking for solutions to the CGL equations that correspond to circularly polarized { plane} waves propagating along the $z$ direction. Those exact solutions will define the dynamical equilibrium for the subsequent linear stability analysis described in section~\ref{results}.  

To this end, consider a uniform magnetized plasma with density $\rho_0$, magnetic field ${\bf B}_{0}=B_{0}{\bf \hat z}$ (say, $B_0>0$) and parallel and perpendicular pressures $p_{0\parallel}$ and $p_{0\bot}$, respectively. {Also, since we are interested in plane waves, the longitudinal velocity and magnetic field $U_z$ and $B_z$ are set to zero.} Transverse velocity and magnetic field  fluctuations, labelled respectively  ${\bf U_\bot}(z,t)$ and ${\bf B_{\bot}}(z,t)$, with constant and homogeneous magnetic pressure {(i.e. $|{\bf B_\bot}|=const.$)} obey to the following  coupled differential equations: 

\begin{equation}
\begin{split}
\frac{\partial {\bf U_{\bot}}}{\partial t}=\frac{B_{0}}{4\pi\rho_0}\frac{\partial}{\partial z}{\bf B_{\bot}}+\frac{1}{\rho_0}\frac{\partial}{\partial z}\left(\frac{ B_{0}{\bf B_{\bot}}}{B_{\bot}^2+B_{0}^2}\Delta p_0\right),
\end{split}
\label{cgl1}
\end{equation}
\begin{equation}
\frac{\partial {\bf B_{\bot}}}{\partial t}=B_{0}\frac{\partial}{\partial z}{\bf U_{\bot}}.
\label{cgl2}
\end{equation}

%%\begin{figure}
%%\includegraphics[width=0.45\textwidth]{FH_th}
%%\caption{Fire-hose threshold $\tilde V_a^2$=0 for different values of the normalized wave amplitude. The region corresponding to propagating waves lies on the left of the plotted curves, while the non-propagating fluctuations lie on the right.}
%%\label{FH}
%%\end{figure}
Note that the above equations remain valid also in cases where the total magnetic pressure is not uniform. In the latter case however, the full complement of the double adiabatic and continuity equations would need to be considered to solve Eqs.~(\ref{cgl1})--(\ref{cgl2}). Here we will always be concerned with the special case in which the total magnetic pressure is uniform in space.

After introducing the normalized magnetic field ${\bf \hat B_{\bot}}={\bf B_{\bot}}/B_{0}$ and the Alfv\'en speed $V_a=B_{0}/\sqrt{4\pi\rho_0}$, we can  combine eqs.~(\ref{cgl1})--(\ref{cgl2}) so as to obtain the following  wave equation:
\begin{equation}
\begin{split}
\frac{\partial^2}{\partial t^2}{\bf \hat B_{\bot}}(z,t)=\left(V_a^2+\frac{1}{\rho_0}\frac{\Delta p_0}{1+\hat B_\bot^2}\right)\frac{\partial^2}{\partial z^2} {\bf \hat B_\bot}(z,t).
\end{split}
\label{cgl3}
\end{equation}

Provided the condition 
\begin{equation}
\tilde V_a^2= V_a^2+\frac{1}{\rho_0}\frac{\Delta p_0}{1+ \hat B_\bot^2}>0
\label{stability}
\end{equation}
is satisfied, Eq.~(\ref{cgl3}) describes Alfv\'en waves, either left or right handed polarized, that propagate at the speed $\tilde V_a$. Forward (upper sign) and backward (lower sign) {propagating} waves have the following general form:  
\begin{equation}
{\bf \hat B_\bot}(z,t)=\hat B_\bot \Re\left[e^{if(z\mp\tilde V_at)}({\bf \hat x}-i{\bf \hat y})\right],
\end{equation}
\begin{equation}
{\bf U_\bot}(z,t)=\mp \tilde V_a {\bf \hat B_\bot}(z,t), 
\end{equation} 
where for the sake of illustration a left-hand polarization has been chosen.

\begin{figure}
\includegraphics[width=0.35\textwidth]{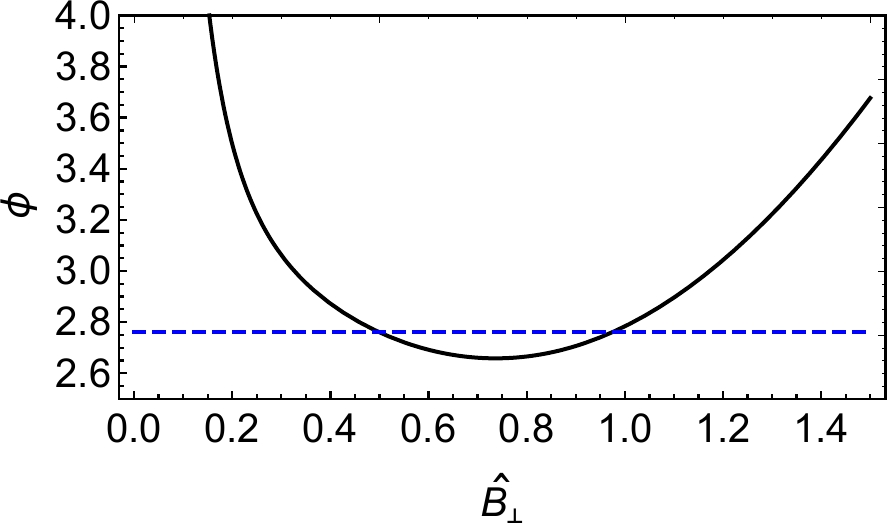}\qquad\qquad\qquad\qquad
\includegraphics[width=0.35\textwidth]{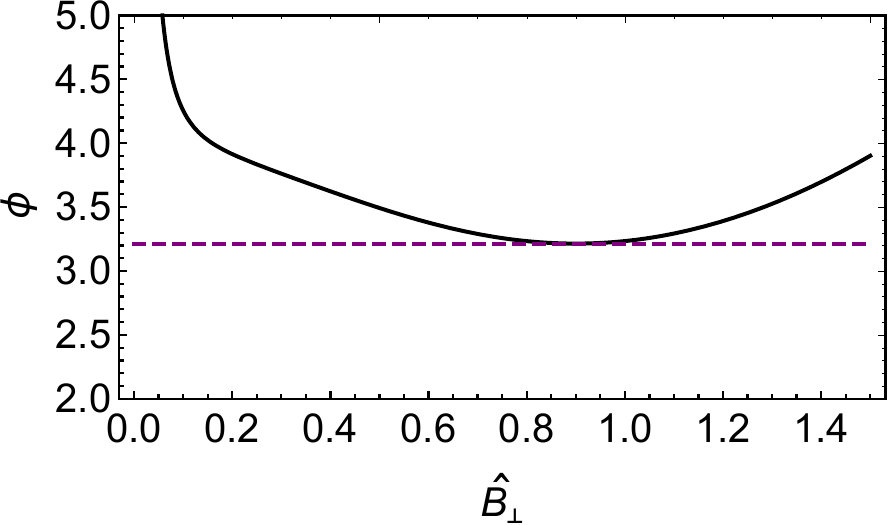}
\caption{Plot of the potential $\phi$  (solid black lines) and of the total energy $E$ (dashed lines) vs. the amplitude $\hat B_\bot$ for $\beta_\parallel=4$ and anisotropic pressure ratio $\xi=0.1$: the upper panel corresponds to a fluctuation in the firehose regime with $\hat B_\bot(0)=0.5$; the lower panel corresponds to a circularly polarized Alfv\'en wave with $\hat B_\bot(0)=0.9$.}
\label{FH2}
\end{figure}

\subsection{General properties of the CGL nonlinear wave equation and ``firehose" fluctuations}
The condition $\tilde V_a^2=0$, with $\tilde V_a^2$ given by Eq.~(\ref{stability}), generalizes the well-known threshold separating Alfv\'en waves from exponentially growing firehose fluctuations by including the nonlinear effect of the fluctuation magnetic pressure -- for a discussion on the firehose instability in the small amplitude limit see e.g. \citet{hunana_2017}. By analogy with the traditional firehose stability condition, the inequality  $\tilde V_a^2>0$ may be expressed in terms of the total magnetic pressure or, in non dimensional form, in terms of the  total parallel plasma beta~$\beta_{\parallel}^{\text{{\sc t}}}$ and $\xi$, defined respectively by
\begin{equation}
\beta_{\parallel}^{\text{{\sc t}}}=8\pi\frac{p_{0\parallel}}{B_0^2}\frac{1}{1+\hat B_\bot^2}\quad \text{and}\quad \xi=\frac{p_{0\bot}}{p_{0\parallel}},
\end{equation}
\begin{equation}
\left( \xi-1 \right)\frac{\beta_{\parallel}^{\text{{\sc t}}}}{2}>-1.
\label{stability2}
\end{equation} 

It is however convenient to leave the amplitude $\hat B_\bot$ as a free parameter and to define $\beta_\parallel$ by means of the background magnetic field,~i.e.
\begin{equation}
\beta_{\parallel}=p_{0\parallel}\frac{8\pi}{B_0^2}.
\label{beta}
\end{equation}

For small amplitude fluctuations the firehose instability is obtained when 
\begin{equation}
 \left(\xi-1\right)\frac{\beta_\parallel}{2}<-1;
\label{th0}
\end{equation}
however, inclusion of finite amplitude effects leads to a critical amplitude $\hat B^*$ above which the condition $\tilde V_a^2>0$ remains satisfied, and therefore waves may still propagate if of sufficiently large amplitude:
\begin{equation}
 \hat B_\bot>\hat B_{\bot}^*=\left[\frac{1}{2}\beta_\parallel \left(1-\xi\right)-1\right]^{1/2}.
\label{th1}
\end{equation}

Consider now  the case of a firehose unstable fluctuation, with a very small  initial amplitude $\hat B_\bot(0)\ll \hat B_\bot^*$. Clearly, the amplitude starts to grow exponentially as predicted by the linear theory, approaching the value of $\hat B_\bot^*$. At that point finite amplitude effects come into  play in a non trivial way preventing the fluctuation from growing without bounds. In order to understand the temporal evolution of fluctuations in this  regime ($\tilde V_a^2<0$) we briefly discuss below the behavior of nonlinear solutions with time-dependent amplitude. 

Nonlinear solutions that generalize circularly polarized Alfv\'en waves must satisfy the following nonlinear wave equation:  
\begin{equation}
\begin{split}
\frac{\partial^2}{\partial t^2}{\bf \hat B_{\bot}}(z,t)=\left(V_a^2+\frac{1}{\rho_0}\frac{\Delta p(t)}{1+\hat B_\bot^2(t)}\right)\frac{\partial^2}{\partial z^2} {\bf \hat B_\bot}(z,t),
\end{split}
\label{nl2}
\end{equation}
where $\Delta p$ evolves in time according to Eqs.~(\ref{CGL_eq1})--(\ref{CGL_eq2}). The latter can be integrated to obtain the following expression:
\begin{equation}
\Delta p(t)=\left[ p_{0\bot}\sqrt{\frac{1+\hat B_\bot^2(t)}{1+ \hat B_\bot^2(0)}}- p_{0\parallel}\frac{1+ \hat B_\bot^2(0)}{1+ \hat B_\bot^2(t)}   \right].
\label{cgl4}
\end{equation}

The boundedness of solutions of Eq.~(\ref{nl2}) can be demonstrated from first principles for a single Fourier mode with wave vector $k_0$  by deriving the corresponding energy conservation equation. In this case fluctuations with homogeneous magnetic pressure  can be written in the following form, 
\begin{equation}
\begin{split}
&\hat B_x(z,t)=\Re[B_{k_0}(t)e^{ik_0z}],\\
&\hat B_y(z,t)=\Im[B_{k_ 0}(t)e^{ik_0z}],
\end{split}
\end{equation}
where $B_{k_0}$ is the complex time dependent amplitude. A standard manipulation of Eq.~(\ref{nl2}) leads to the  total energy ($E$) conservation equation, involving the square amplitude of the fluctuation ($\hat B_\bot^2=\hat B_x^2+\hat B_y^2$),  its square temporal derivative (denoted with a dot, $\dot B_\bot^2=(\dot B_x)^2+(\dot B_y)^2$) and the effective potential ($\phi$): 
\begin{equation}
\dot B_\bot^2+\phi=E. 
\label{cgl5}
\end{equation}
By defining $\omega_A^2= k_0^2 V_a^2$   and $L=2\hat B_\bot^2\dot\theta$, where  $L$  is a constant determined by the initial conditions,  $\theta$ being the angle between $\hat B_y$ and $\hat B_x$, the potential $\phi$ can be written in compact form as
\begin{equation}
\begin{split}
 \phi=&\omega_A^2 \hat B_\bot^2+\frac{1}{4}\frac{L^2}{\hat B_\bot^2}+\\
 &\omega_A^2 \frac{\beta_\parallel}{2}\left(2\xi\frac{\sqrt{1+\hat B_\bot^2}}{\sqrt{1+\hat B_\bot^2(0)}}+\frac{1+\hat B_\bot^2(0)}{1+\hat B_\bot^2}\right).
\end{split}
\label{cgl6}
\end{equation}

Equations~(\ref{cgl5})--(\ref{cgl6}) show that the system is equivalent to the motion of a particle subject to a central field of force, where the perpendicular amplitude and its temporal derivative play the role of the particle position and velocity, respectively,  $L$ is the particle angular momentum and $\phi$ is the potential energy which is a function of coordinates. The first term on the right hand side of Eq.~(\ref{cgl6}) corresponds to the harmonic potential, the second term is the counterpart of the centrifugal potential  and the last term is introduced by the double adiabatic closure. At large amplitudes the harmonic potential dominates, and therefore $\phi$ behaves as a potential well for any $\beta_\parallel$ and $\xi$. As a consequence, there are no unbounded solutions for firehose unstable fluctuations: minima and maxima of the wave amplitude are the turning points corresponding to $E=\phi$, whereas constant amplitude waves (i.e. circularly polarized waves) lie in the potential minimum. 

\begin{figure}[tb]
\includegraphics[width=0.35\textwidth]{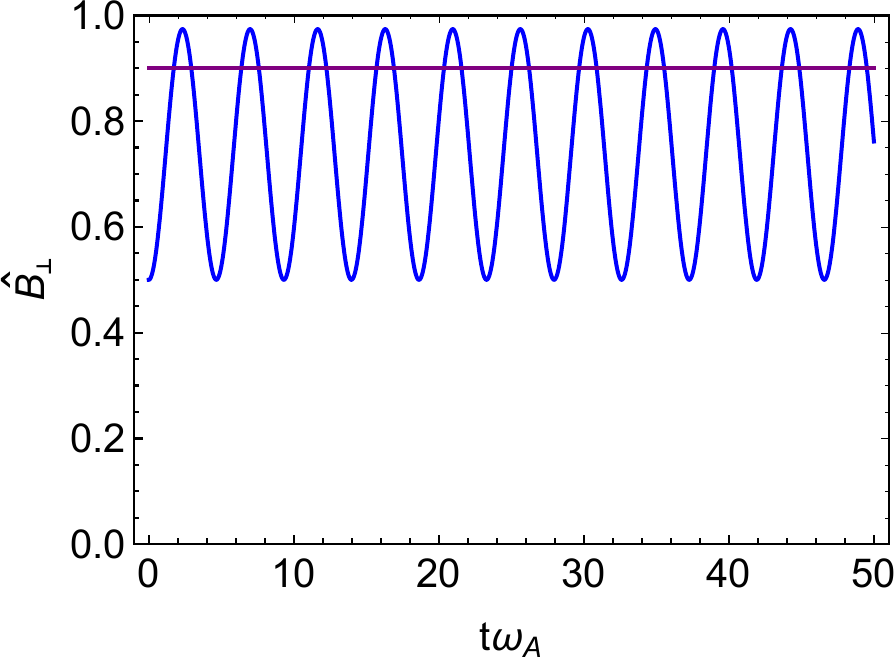}
\caption{Temporal evolution of the amplitude of a circularly polarized Alfv\'en wave (purple color) and of a non-constant amplitude fluctuation in the firehose regime (blue color), for the same parameters given in Fig.~\ref{FH2}.}
\label{FH3a}
\end{figure}

As an example, we show in Fig.~\ref{FH2} the plot of $\phi$ for $\xi=0.1$ and $\beta_\parallel=4$, therefore for the more interesting case that can be in the firehose regime depending upon the initial amplitude. Dashed lines correspond to the total energy $E$ fixed by the initial conditions, that we show for two cases: for $\hat B_\bot(0)=0.5$, smaller than $\hat B_\bot^*$ (upper panel), and  for $\hat B_\bot(0)=0.9$, larger than $\hat B_\bot^*$ (lower panel). In Fig.~\ref{FH3a} and \ref{FH3b} we plot the magnitude of $\hat B_\bot(t)$ as a function of time and its polarization at a fixed point in space, respectively, for the same two cases indicated in Fig.~{\ref{FH2}. The plotted solutions were obtained by integrating numerically Eq.~(\ref{nl2}) for a monochromatic wave and by imposing initial conditions that  correspond to a forward circularly polarized wave: $\hat B_x(0)=\hat B_\bot(0)$, $\hat B_y(0)=0$ and ${\dot B_x(0)=0}$, ${\dot B_y(0)}=-\omega_A(|\tilde V_a(0)|/V_a)\hat B_\bot(0)$. As expected, for $\hat B_\bot(0)>\hat B_\bot^*$ (purple color in Fig.~\ref{FH3a} and \ref{FH3b}) the amplitude of the wave remains constant and has minimum potential energy (see Fig.~\ref{FH2}, lower panel), corresponding to a circularly polarized wave; in the case $\hat B_\bot(0)<\hat B_\bot^*$ (blue color in Fig.~\ref{FH3a} and \ref{FH3b}), the fluctuation is initially in the firehose regime and its amplitude oscillates between the two turning points $\hat B_\bot=0.5$ and $\hat B_\bot=0.98$ (see Fig.~\ref{FH2}, upper panel). 

In conclusion, in the framework of the CGL model the firehose threshold does not represent a stability condition for fluctuations with spatially homogeneous (total) pressure, but rather it represents a condition for the existence of time-constant amplitude waves{. Below that threshold the wave amplitude has to be an evolving function of time, so that the condition $|{\bf B_\bot}|=const$ cannot be satisfied}.  Eq.~(\ref{th1})  therefore provides a lower limit for the amplitude of Alfv\'en waves below which circularly polarized Alfv\'en waves cannot exist, that at large $\beta_\parallel$ roughly scales~as 
\begin{equation}
\hat B_\bot^*\sim\beta_\parallel^{1/2}.
\label{scaling}
\end{equation}

It is worthwhile to notice that the scaling given in Eq.~(\ref{scaling}) has an opposite trend with $\beta_\parallel$ with respect to the one found by ~\cite{squire}, who studied finite amplitude effects on the propagation of linearly polarized Alfv\'en waves at very large $\beta_\parallel$. They found an amplitude upper limit $\hat B_\bot^{max}$ scaling as  $\hat B_\bot^{max}\sim\beta^{-1/2}$ ($\beta$ defined with an average pressure $p_0=2/3p_{0\bot}+1/3p_{0\parallel}$), above which Alfv\'en  waves  become strongly modified by a { self-}induced firehose instability: the temporal and spatial modulations of the total magnetic field magnitude lead to variations of $\Delta p$ that drives the system towards the firehose unstable threshold thereby ``interrupting'' the wave.  The latter evolves into a sequence of spatial discontinuities so as to  minimize $\hat B_\bot^2$ due to a third order effect in the wave amplitude. This mechanism  does not take place in our case, where we consider fluctuations with  a constant magnitude of the magnetic field. However, we conjecture that at large $\beta_\parallel$ any perturbation above a circularly polarized state will return the field to a configuration of total constant magnetic field magnitude rather than the ``interrupted" profiles seen starting from linear polarization. Indeed we believe this process may be at work in the formation of the constant amplitude Alfv\'en waves seen in the solar wind, a question that will be explored in a subsequent paper. Such a mechanism cannot  be captured in the present study where we retain up to first order terms in the perturbation amplitude.

\begin{figure}[tb]
\includegraphics[width=0.35\textwidth]{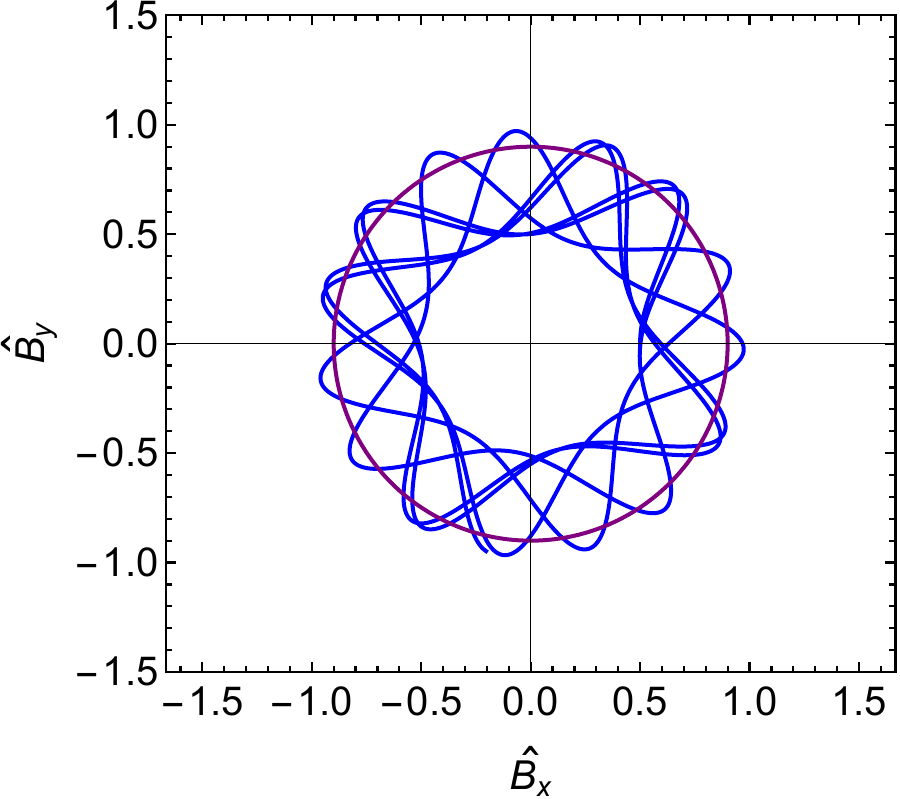}
\caption{Polarization parameterized in time for the circularly polarized wave and the fluctuation in the firehose regime described in Fig.~\ref{FH2}--\ref{FH3a}.}
\label{FH3b}
\end{figure}

\section{Parametric instability: results}  
\label{results}

We now take a monochromatic, left-hand circularly polarized mother Alfv\'en wave in parallel propagation with magnetic and velocity field as given below,
\begin{equation}
{\bf \hat B_{\bot}}=\hat B_\bot e^{i(k_0z-\omega_0t)}{\bf(\hat x}-i{\bf \hat y)}, 
\label{aw1}
\end{equation} 
\begin{equation}
 {\bf  U_{\bot}}=-\tilde V_a{\bf \hat B_{\bot}},
 \label{aw2}
\end{equation} 
\begin{equation}
\omega_0=\tilde V_a k_0,
 \label{aw3}
\end{equation} 
and, following the standard method, we study its stability with respect to small perturbations in velocity, pressure, density and magnetic field. Perturbations are of the form 
\begin{equation}
u_\parallel e^{i(kz-\omega t)}, \quad {u_\pm} e^{i(k\pm k_0)z-i(\omega\pm\omega_0) t},
\end{equation} 
\begin{equation}
p_{\parallel(\bot)} e^{i(kz-\omega t)}, \quad \rho e^{i(kz-\omega t)},
\end{equation} 
\begin{equation}
 { \hat b_\pm} e^{i(k\pm k_0)z-i(\omega\pm\omega_0) t},
\end{equation} 
where $u_\pm=u_y\pm i u_z$ and similarly  $\hat b_\pm=\hat b_y\pm i \hat b_z$. Linearization of the parent system of Eqs.~(\ref{CGL_eq0})--(\ref{CGL_eq2}) around the configuration given in Eqs.~(\ref{aw1})--(\ref{aw2}), leads to the following set of algebraic equations,

%\begin{widetext}
\begin{equation}
\omega\frac{\rho}{\rho_0}-ku_\parallel=0
\label{linearCGL1}
\end{equation}
\begin{equation}
(\omega\pm\omega_0)\hat b_\pm+(u_\pm-\hat B_\bot u_\parallel)(k\pm k_0)=0
\end{equation}
\begin{equation}
\begin{split}
\omega u_\parallel=k\frac{p_\bot}{\rho_0}+\hat B_\bot k(\hat b_++\hat b_-)&\left(\frac{1}{2}V_a^2+\frac{\Delta p_0}{\rho_0(1+\hat B_\bot^2 )^2}  \right)\\
&-k\frac{\Delta p}{\rho_0(1+\hat B_\bot^2 )}
\end{split}
\end{equation}
\begin{equation}
\begin{split}
(\omega\pm\omega_0) u_\pm&\pm \frac{\rho}{\rho_0}\omega_0u_{0\pm}\mp k_0u_{0\pm}u_\parallel=-V_a^2(k\pm k_0)\hat b_\pm\\
&-(k\pm k_0)\left[\frac{\Delta p_0}{\rho_0(1+\hat B_\bot^2 )}\hat b_\pm +\frac{\hat B_\bot }{1+\hat B_\bot^2 }\frac{\Delta p}{\rho_0}\right. \\
&\left. -\frac{\hat B_\bot^2 }{(1+\hat B_\bot^2 )^2}\frac{\Delta p_0}{\rho_0}(\hat b_++\hat b_-)  \right]
\end{split}
\end{equation}
\begin{equation}
p_\parallel=p_{0\parallel}\left[  3\frac{\rho}{\rho_0}-\frac{\hat B_\bot }{1+\hat B_\bot^2 }(\hat b_++\hat b_-) \right]
\end{equation}
\begin{equation}
p_\bot=p_{0\bot}\left[  \frac{\rho}{\rho_0}+\frac{1}{2}\frac{\hat B_\bot }{1+\hat B_\bot^2 }(\hat b_++\hat b_-) \right]
\end{equation}
\begin{equation}
\begin{split}
\Delta p=&\frac{\rho}{\rho_0}(p_{0\bot}-3p_{0\parallel})+\\
&\frac{\hat B_\bot }{1+\hat B_\bot^2 }(\hat b_++\hat b_-) (\frac{1}{2}p_{0\bot}+p_{0\parallel}).
\label{linearCGL2}
\end{split}
\end{equation}
%\end{widetext}

The dispersion relation for the complex frequencies relative to Eqs.~(\ref{linearCGL1})--(\ref{linearCGL2}) can be written in implicit form as follows: 
\begin{widetext}
\begin{equation}
\begin{split}
&\left[ \hat \omega^2-\tilde \beta \hat k^2\left( 1+\frac{\hat B_\bot^2\xi}{3} \right)\right]\left\{(\hat \omega-\hat k)[(\hat \omega+\hat k)^2-4]+\frac{\tilde\beta\hat B_\bot^2(\xi-4)}{3(1+\hat B_\bot^2)} [(\hat k^2+1)\hat \omega+\hat k(\hat k^2-3)] \right\}\\
&=\hat B_\bot^2 \hat k^2\left[  1-\frac{\tilde\beta(3-\xi-\hat B_\bot^2)}{3(1+\hat B_\bot^2)}\right]\{(\hat \omega^3+\hat \omega^2\hat k-3\hat \omega+\hat k)-\frac{\tilde\beta (3-\xi)}{3}[(\hat k^2+1)\hat \omega+\hat k(\hat k^2-3)]\}.
\end{split}
\label{dr}
\end{equation}
\end{widetext}

Equation~(\ref{dr}) is expressed in normalized units with speeds  normalized to $\tilde V_a$ and frequencies to $\omega_0$. For the sake of notation we have defined 

\begin{equation}
 \hat \omega=\omega/\omega_0,\quad \hat k=k/k_0,
\end{equation}
and  the parameter $\tilde\beta$ is defined~as
\begin{equation}
\tilde\beta=\frac{3}{2}\frac{\beta_\parallel}{1+\hat B_\bot^2+\frac{\beta_\parallel}{2}(\xi-1)}.
\end{equation}
\begin{figure}[htb]
\includegraphics[width=0.35\textwidth]{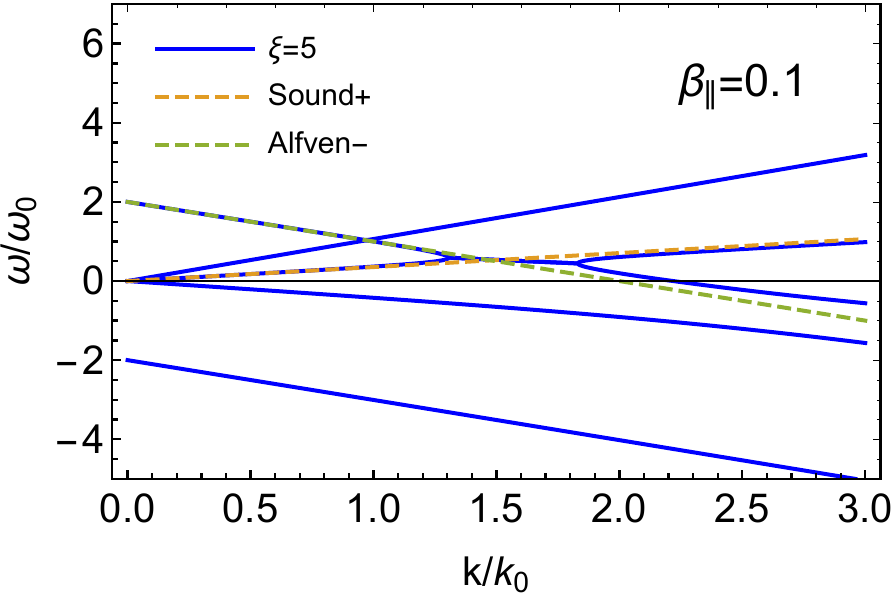}
\includegraphics[width=0.35\textwidth]{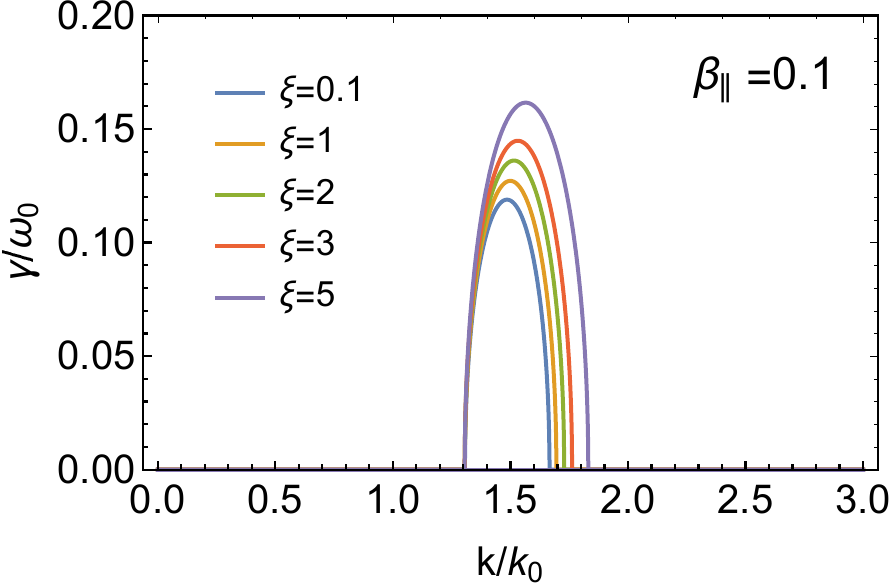}
\caption{Normalized frequency $\omega/\omega_0$ and growth rate $\gamma/\omega_0$ vs. $k/k_0$ for $\beta_\parallel=0.1$ and $\hat B_\bot^2=0.1$ at different anisotropic pressure ratio~$\xi$.}
\label{ex1}
\end{figure}
\begin{figure}[htb]
\includegraphics[width=0.35\textwidth]{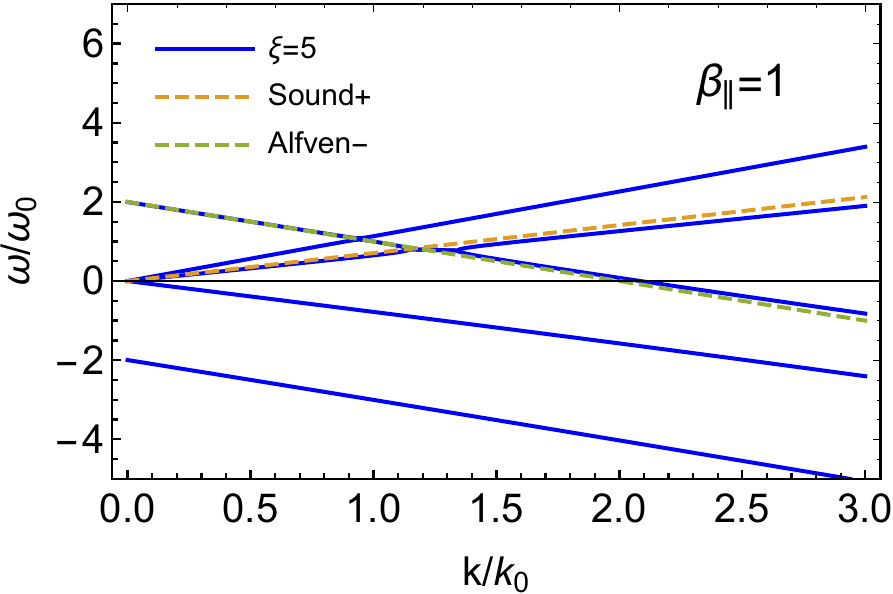}
\includegraphics[width=0.35\textwidth]{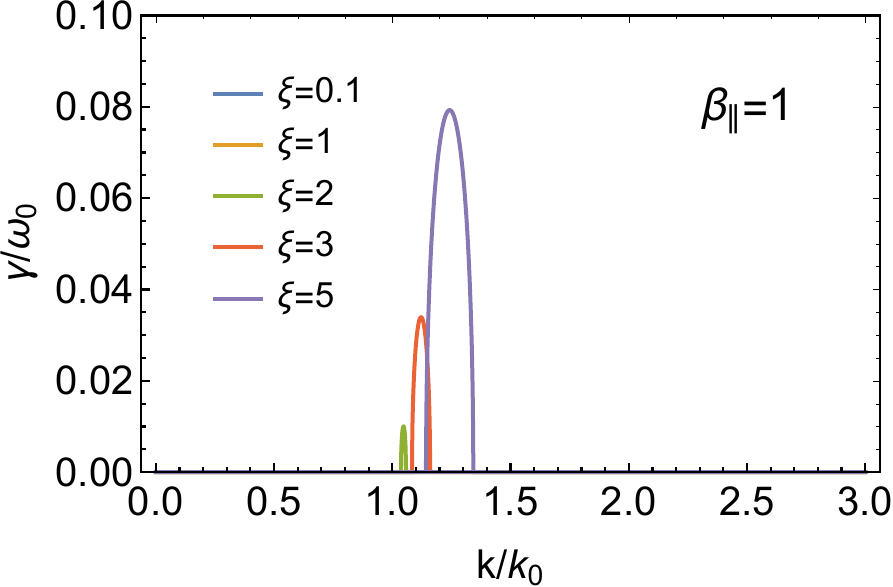}
\caption{Normalized frequency $\omega/\omega_0$ and growth rate $\gamma/\omega_0$ vs. $k/k_0$ for $\beta_\parallel=1$ and $\hat B_\bot^2=0.1$ at different anisotropic pressure ratio~$\xi$.}
\label{ex2}
\end{figure}

The dispersion relation given in Eq.~(\ref{dr}) is a fifth order equation in $\hat\omega$ that we have solved numerically for different values of the plasma parameters $\hat B_\bot^2$,  $\xi$ and $\beta_\parallel$. We therefore prefer to use  the plasma beta $\beta_\parallel$ defined with the magnitude of the average magnetic field $B_0$ (see Eq.~\ref{beta}) instead of $\beta_\parallel^{\text{\sc t}}$ or $\tilde\beta$,  so that all plasma parameters are independent from each other. 

In Fig.~\ref{ex1} and \ref{ex2}  we show the real  part  of the dispersion relation ($\omega(k)$, upper panel) for anisotropic pressure ratio $\xi=5$ and the imaginary part ($\gamma(k)$, lower panel) for different values of $\xi$ at fixed $\hat B_\bot^2=0.1$ and, respectively,  $\beta_\parallel=0.1$ and $\beta_\parallel=1$.  For clarity, we also display in dashed lines the dispersion relation of the two main interacting branches for $\hat B_\bot^2=0$: the sound branch $\omega=\tilde V_a\tilde\beta^{1/2}k$ in orange and the backward Alfv\'en branch $\omega=-\tilde V_a(k-k_0)+\omega_0$ in green. As can be seen by inspection, there is a range of  wave vectors unstable to the parametric decay due to the coupling between  a forward sound wave  with $ k>k_0$  and a low frequency backward right-handed  Alfv\'en wave, with wave vector $ k_-= k-k_0$ and frequency $\omega_-=\omega-\omega_0$. At larger $\beta_\parallel$ the interaction becomes a four-wave interaction including also a forward Alfv\'en wave.  Fig.~\ref{ex1} and \ref{ex2} (bottom panels) show that in general the range of unstable modes and the growth rate increase with $\xi$  and that, provided $\xi$ is large enough, parametric decay can occur for relatively small amplitudes (e.g. $\hat B_\bot^2=0.1$) even at large plasma beta (e.g. $\beta_\parallel\simeq1$), contrary to the MHD model.

\begin{figure*}[t]
\includegraphics[width=0.33\textwidth]{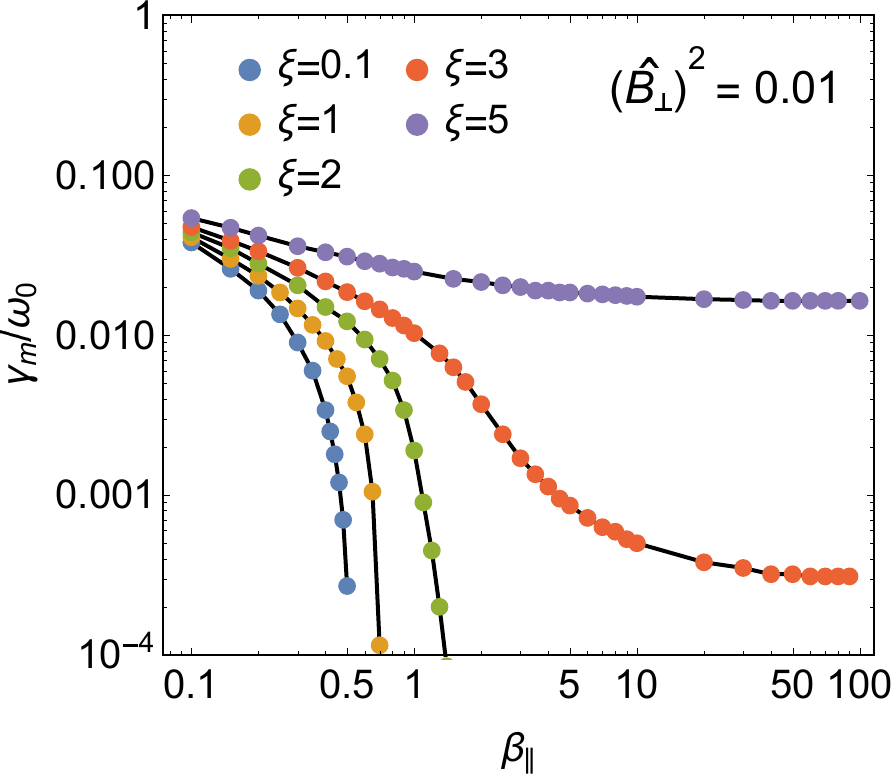}
\includegraphics[width=0.33\textwidth]{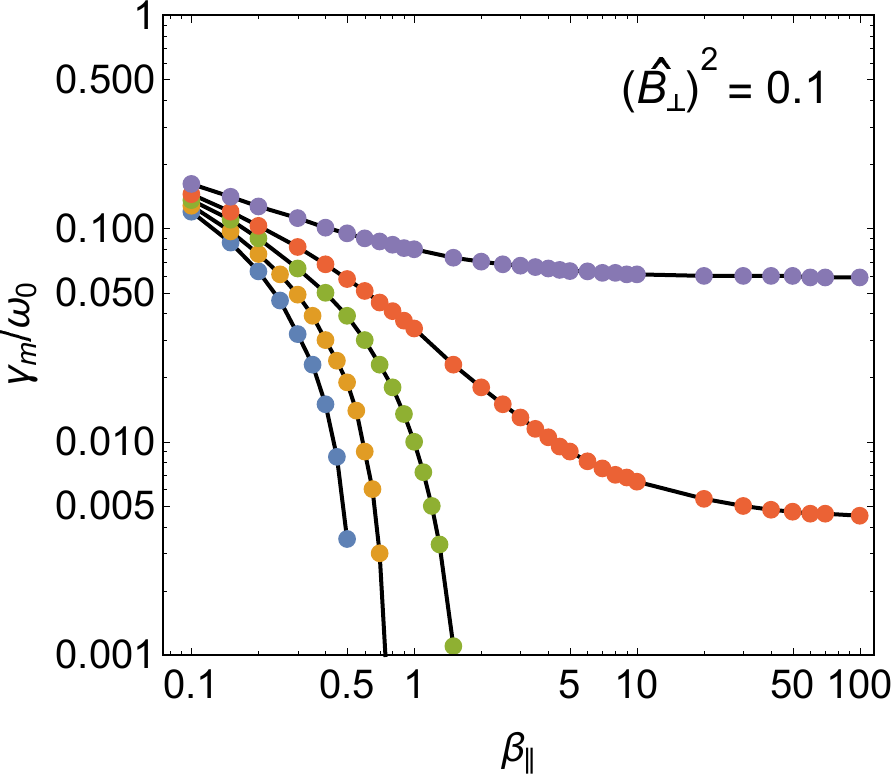}
\includegraphics[width=0.33\textwidth]{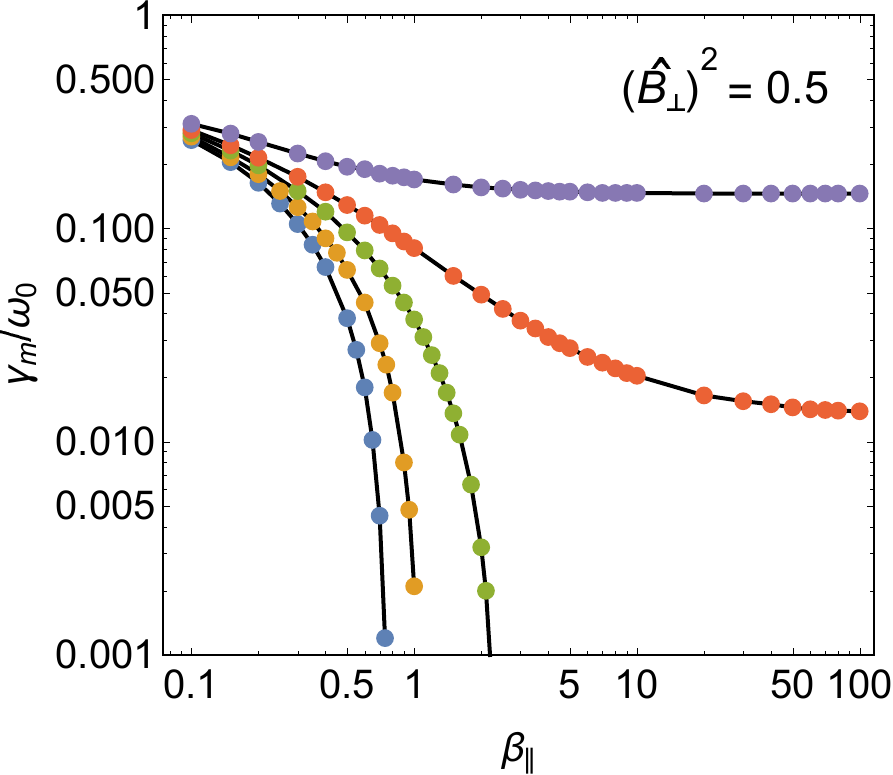}
\caption{Maximum growth rates $\gamma_m/\omega_0$ as a function of $\beta_\parallel$ at different anisotropic pressure ratio $\xi$, for $\hat B_\bot^2=0.01$  (left panel),  $\hat B_\bot^2=0.1$  (middle panel) and $\hat B_\bot^2=0.5$  (right panel).}
\label{ex3}
\end{figure*}

The background pressure anisotropy  affects the properties of the instability in different ways depending upon the value of~$\beta_\parallel$. In the limit $\beta_\parallel\rightarrow0$ the dispersion relation given in Eq.~(\ref{dr}) is independent from both $\beta_\parallel$ and $\xi$, and  the known MHD results are therefore recovered for small values of $\beta_\parallel$\footnote{The limit $\beta_\parallel\rightarrow0$ of Eq.~(\ref{dr}) is: $\hat \omega^2(\hat \omega-\hat k)[(\hat \omega+\hat k)^2-4]=\hat B_\bot^2 \hat k^2(\hat \omega^3+\hat \omega^2\hat k-3\hat \omega+\hat k)$. This corresponds to the MHD dispersion relation of the parametric decay in a cold plasma~\citep{derby}.}. Interestingly, growth rates are independent from $\beta_\parallel$ also in the opposite limit $\beta_\parallel\rightarrow\infty$, but in this case growth rates still depend on  $\xi$~\footnote{{A way to see this is by considering that if $\beta\rightarrow\infty$ then $\tilde\beta\rightarrow 3/(\xi-1)$.}}. Numerical inspection of Eq.~(\ref{dr}) within this limit (where only the case $\xi\geq1$ is considered for consistency with the firehose threshold) shows that, at given amplitude, growth rates are greater than zero if the anisotropic pressure ratio is above a critical value  $\xi^*\simeq2.67$. This implies that the parametric decay instability has growth rates $\gamma(k)$ that decrease and tend to zero as $\beta_\parallel$ increases (unless the amplitude of the mother wave increases with $\beta_\parallel$) if $\xi<\xi^*$, while  it reaches finite and constant $\gamma(k)$ for $\beta_\parallel\gg1$ if $\xi\geq\xi^*$. 

The existence of these two regimes at large $\beta_\parallel$ is confirmed by Fig.~\ref{ex3}, where we show the normalized maximum growth rate $\gamma_m/\omega_0$ as a function of $\beta_\parallel$ for different values of $\xi$ and $\hat B_\bot^2$. As can be seen by inspection of the displayed plots,  if $\xi<\xi^*$ the maximum growth rate decreases as $\beta_\parallel$ increases, similarly to what happens in the MHD model. On the contrary, for $\xi>\xi^*$ the maximum growth rate becomes  weakly dependent on $\beta_\parallel$ as the latter increases, and it tends to a constant value for $\beta_\parallel\gg1$. At given amplitude, all curves $\gamma_m$ converge towards the same value for $\beta_\parallel\ll1$.

Another useful way to display our results is to plot the contours of the maximum growth rates in parameter space.  In Fig.~\ref{contours} and \ref{contours2} we show  the contours corresponding to $\gamma_m/\omega_0=0.01,\,0.05,\,0.1$ (blue  lines) together with the contour line corresponding to $\tilde V_a^2=0$ (red  line). The region above of the red line is unstable to  parametric decay, while the region below the red line has not been considered in this study. 

In Fig.~\ref{contours} we display the contours in the $(\xi,\,\beta_\parallel)$ plane for three fixed square amplitudes of the mother wave, $\hat B_\bot^2=0.5$ (left panel), $\hat B_\bot^2=0.5$ and $\hat B_\bot^2=1$ (right panel). As can be seen, growth rates do not depend on the plasma beta at large $\beta_\parallel$, whereas for small $\beta_\parallel$ they essentially are unaffected by the anisotropy. As the amplitude increases, contours display the same trends but they are shifted towards larger $\beta_\parallel$ and smaller $\xi$ values. The firehose threshold is also shifted, in this case due to the fact that larger amplitudes have a higher threshold  (cfr. Eq.~\ref{th1}). In Fig.~\ref{contours2} the same growth rates are shown in the $(\hat B_\bot,\,\beta_\parallel)$ plane for $\xi=5$, larger than $\xi^*$ (left panel), $\xi=1$ (middle panel) and  $\xi=0.1$, smaller than $\xi^*$ (right panel). These plots also clearly show the existence of two regimes: if $\xi>\xi^*$ then the growth rates become independent on $\beta_\parallel$ and hence the amplitudes saturate at $\beta_\parallel\gg1$ (Fig.~\ref{contours2} left panel); if instead $\xi<\xi^*$, then the trends are similar to those found in MHD, with the parametric instability being strongly stabilized as $\beta_\parallel$ increases. In this regime of anisotropy amplitudes can be extremely large. In particular, amplitudes  scale as $\hat B_\bot\sim \beta_\parallel^{1/2}$, for $\beta_\parallel\gg1$ and fixed $\gamma_m$,  like the firehose threshold (Fig.~\ref{contours2}, middle and right panels). 

\begin{figure*}[htb]
\includegraphics[width=0.33\textwidth]{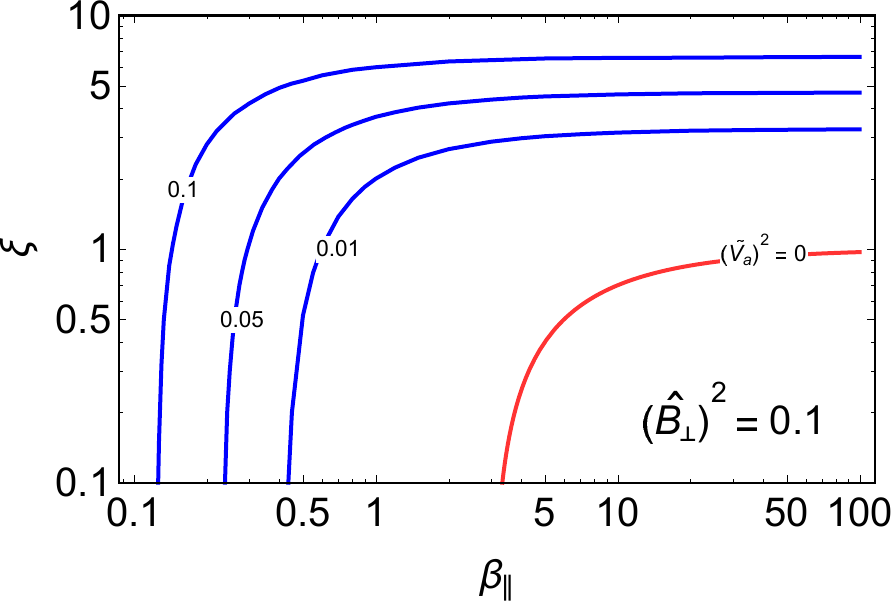}
\includegraphics[width=0.33\textwidth]{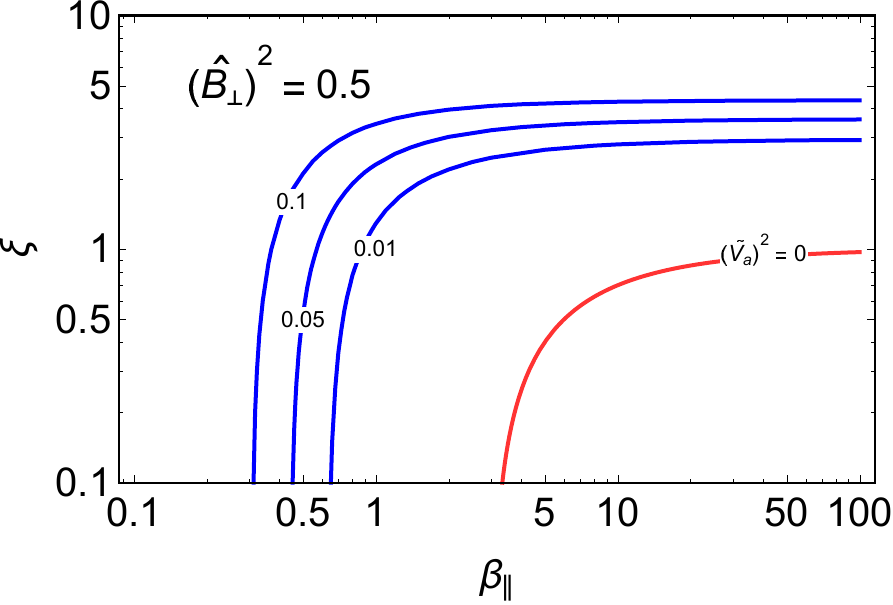}
\includegraphics[width=0.33\textwidth]{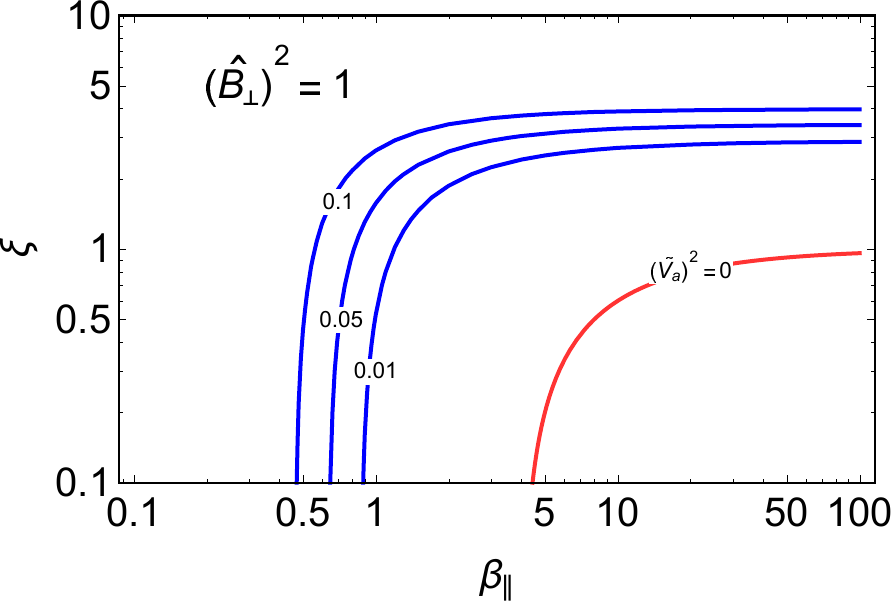}
\caption{Contours of the normalized maximum growth rate $\gamma_m/\omega_0=0.01,\,0.05,\,0.1$ in the parameter space $(\xi,\,\beta_\parallel)$ (blue  lines) and the threshold given by Eq.~(\ref{th1}) (red line),  for $\hat B_\bot^2=0.1$ ({ left} panel), $\hat B_\bot^2=0.5$ ({ middle} panel) and $\hat B_\bot^2=1$ ({ right} panel). }
\label{contours}
\end{figure*}
\begin{figure*}[htb]
\includegraphics[width=0.33\textwidth]{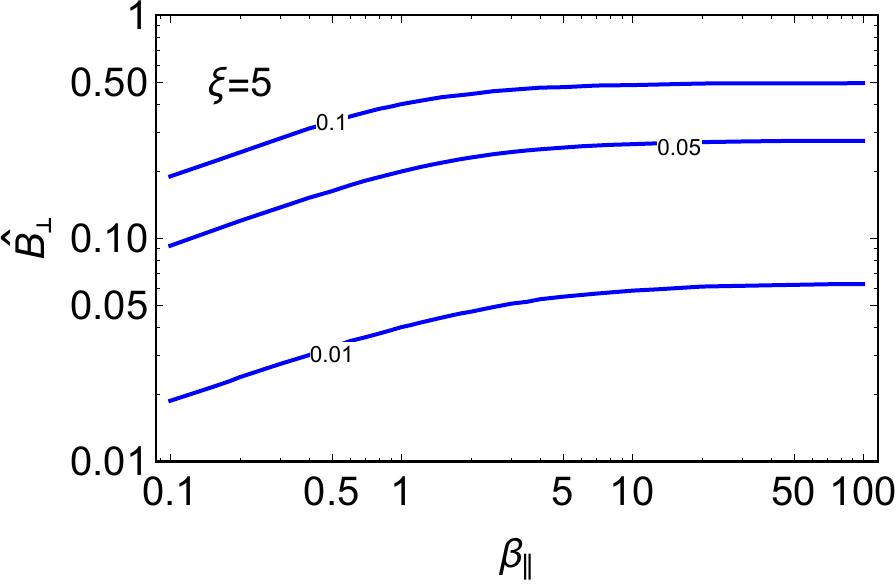}
\includegraphics[width=0.33\textwidth]{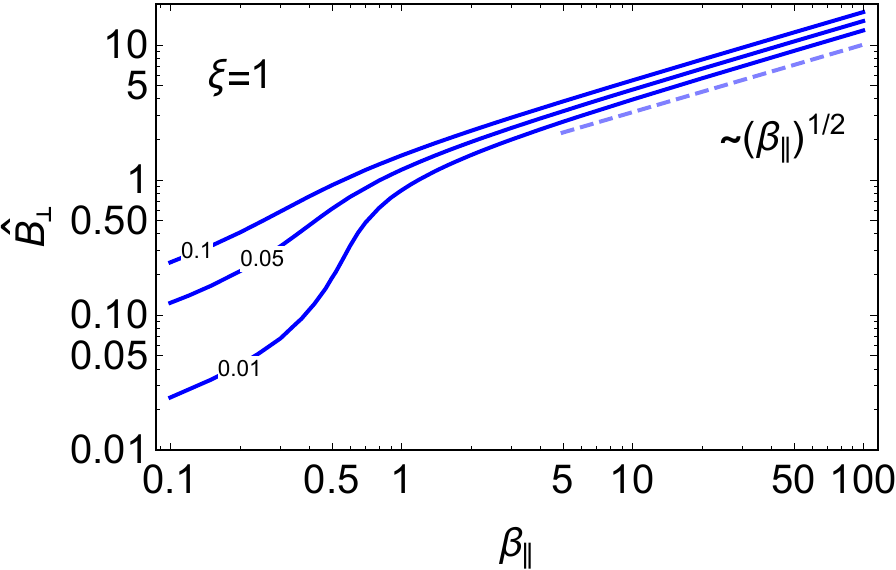}
\includegraphics[width=0.33\textwidth]{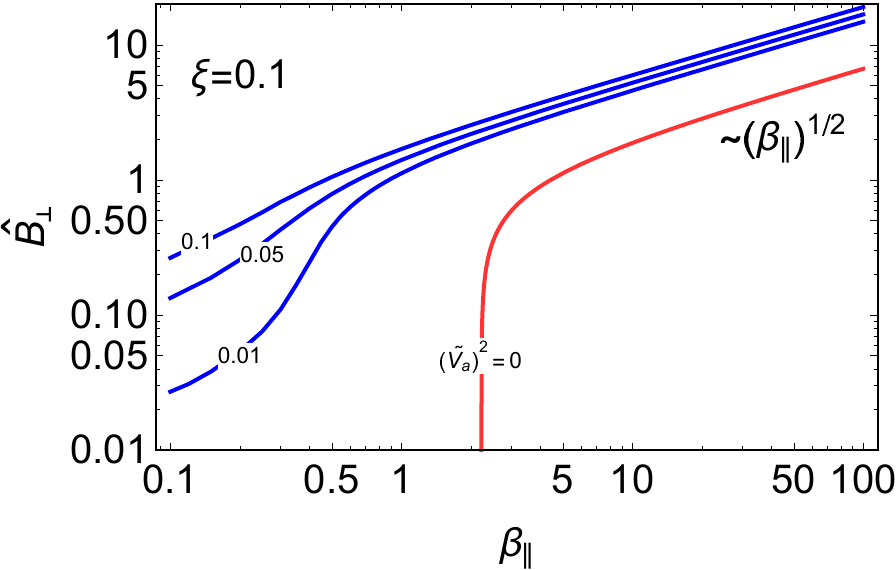}
\caption{Contours of the normalized maximum growth rate $\gamma_m/\omega_0=0.01,\,0.05,\,0.1$ in the parameter space $(\hat B_\bot,\,\beta_\parallel)$ (blue  lines) for $\xi=5$ (left panel), $\xi=1$ (middle panel) and $\xi=0.1$ (right panel). The red line corresponds to the threshold given in  Eq.~(\ref{th1}). At large $\beta_\parallel$ the amplitude of a parametrically unstable Alfv\'en wave scales as $\sim \beta_\parallel^{1/2}$, at given $\gamma_m$.}
\label{contours2}
\end{figure*}

\section{Discussion: the parametric  instability in the solar wind}
\label{disc}

In this section we compare our results with solar wind data. Previous numerical work using the MHD (accelerating) expanding box has shown that the solar wind expansion stabilizes parametric decay unless the growth rate is larger than the expansion rate $\tau_e^{-1}\simeq U_{\text{\sc sw}}/r$, where $U_{\text{\sc sw}}$ is the solar wind speed and $r$ is the heliocentric distance. The condition $\gamma_m\tau_e\simeq1$ can therefore be used now as a criterion for the onset of  parametric decay in the solar wind.  By fixing $U_{\text{\sc sw}}=700$~km/s, we estimate that for low frequencies $\omega_0^{\text{\sc l}}\simeq  2\pi\times10^{-4}$~Rad/s and  high frequencies $\omega_0^{\text{\sc h}}\simeq 2\pi\times10^{-2}$~Rad/s,  the normalized expansion rates are respectively $1/(\tau_e\omega_0^{\text{\sc l}})\simeq10^{-2}$ and $1/(\tau_e\omega_0^{\text{\sc h}})\simeq10^{-4}$, at $r=0.9$~AU, and $1/(\tau_e\omega_0^{\text{\sc l}})\simeq 2\times10^{-2}$ and $1/(\tau_e\omega_0^{\text{\sc h}})\simeq 2\times10^{-4}$ at $r= 0.3$~AU. 

Fig.~\ref{contours_SW} displays diagrams in the parameter space $(\beta_\parallel,\xi)$ of the stable and unstable regions to parametric decay at $r=0.9$~AU (upper panel) and $r=0.3$~AU (lower panel): blue lines correspond to the contour $\gamma_m\tau_e=1$ for $\omega_0^{\text{\sc l}}$ (solid line) and for $\omega_0^{\text{\sc h}}$ (dashed line), whereas the red dot indicates the location  in the diagram of the average fast solar wind. In order to plot the contours, we estimated the normalized square amplitudes $(\hat B_\bot^{\text{\sc l,h}})^2$ that correspond to $\omega_0^{\text{\sc l,h}}$ from the spectra measured by the Helios spacecraft in the Alfv\'enic wind (see e.g. \citet{carbone}): { $(\hat B_\bot^{\text{\sc l}})^2\simeq0.1$  and $(\hat B_\bot^{\text{\sc h}})^2\simeq0.01$ at $r=0.9$~AU, and $(\hat B_\bot^{\text{\sc l}})^2\simeq0.06$ and $(\hat B_\bot^{\text{\sc h}})^2\simeq0.05$  at $r=0.3$~AU.}  To estimate the average values of $\beta_\parallel(r)$, $\xi(r)$ and $B_0(r)$ we used the fast wind empirical profiles discussed in \citet{hellinger_2011}.  Since the low frequency fluctuations are stabilized by the expansion more than the high frequency ones, their threshold provides an upper limit for the stability of a broader spectrum of fluctuations, above which all frequencies are expected to be unstable (white region: $\gamma\tau_e>1$ for $\omega_0>\omega_0^{\text{\sc l}}$). Viceversa, below the threshold  for the highest frequencies the whole spectrum is stable (light blue  region: $\gamma\tau_e<1$ for  $\omega_0<\omega_0^{\text{\sc h}}$). In the yellow region instead the intermediate frequencies $\omega_0^{\text{\sc l}}<\omega_0<\omega_0^{\text{\sc h}}$ are unstable. 

The diagrams displayed in Fig.~\ref{contours_SW} show that the average fast solar wind {is close to the stable region} at $r=0.9$~AU, while at $r=0.3$~AU it moves towards the unstable one. Of course, these diagrams should be regarded as indicative, since slightly different estimations for the expansion rates and amplitudes cause the contour lines to shift somewhat. In addition, fast  solar wind is scattered in the $(\beta_\parallel,\,\xi)$ space, with values of $\xi$ and $\beta_\parallel$  in the range $0.4\lesssim\xi\lesssim3$ and $0.1\lesssim\beta_\parallel\lesssim3$, respectively (see e.g. \citet{matteini_JGR_2013}). The distribution of solar wind  data therefore displays tails that would extend in both the stable and unstable regions. These uncertainties however do not affect the {trend} that high frequency Alfv\'enic fluctuations ($f\simeq10^{-3}-10^{-2}$~Hz) are expected to decay, {and that the region of instability in parameter space broadens while approaching the Sun, extending in this way the range of unstable modes towards the lower frequencies}.  Although it remains to be investigated where  Alfv\'enic periods are located in such diagrams, we argue that signatures of parametric decay -- such as increased content of compressive and inwards Alfv\'en modes, and perhaps a steep radial decrease of amplitudes -- should be more evident at heliocentric distances around $r=0.3$~AU and below, soon to be explored by the upcoming mission Parker Solar Probe~\citep{fox}.  

\section {Summary}
\label{sum}
We have discussed under which conditions a monochromatic,
circularly polarized Alfv\'en wave constitutes an exact
nonlinear state in an anisotropic plasma and we have
studied its stability to parametric decay within the CGL
framework. We have found that in general the growth
rates and the range of unstable modes of the parametric
instability increase with the ratio $p_{0\bot}/p_{0\parallel}$, and that
the background anisotropy introduces a new unstable
regime: for $p_{0\bot}/p_{0\parallel}$ less than a critical value $\xi^*\simeq2.67$ the decay is strongly suppressed at increasing values of $\beta_\parallel$ -- unless the amplitude of the mother Alfv\'en wave
scales as $\beta_\parallel^{1/2}$ -- and it is very similar to parametric decay
in MHD; for $p_{0\bot}/p_{0\parallel}$ {larger than $\xi^*$}, parametric decay occurs
for any $\beta_\parallel$ and, in the limit $\beta_\parallel\gg1$, it becomes independent
of $\beta_\parallel$.  A comparison with typical values of solar
wind plasma beta, anisotropy and amplitudes shows that at $r=0.9$~AU 
the {solar wind is close to the stable region, while at $r=0.3$~AU it moves towards the unstable region}. Decay signatures
should therefore be more evident at heliocentric distances below $r = 0.3$~AU. { Since this study relies on the monochromatic assumption, it will be of interest to inspect in future works whether and how the present results change in the case of more general non-monochromatic fluctuations~\citep{cohen, malara_Phys_fluids_1996}, also by means of numerical simulations in the expanding solar wind generalizing the results of~\cite{anna1}.} 

{Finally, we mention that temperature anisotropy-driven kinetic
instabilities may develop for the same values of plasma beta and thermal pressure anisotropy considered
for parametric decay, namely, the firehose and the ion-cyclotron instability. 
These instabilities} amplify waves in general
at small (kinetic) scales, beyond the validity of the
CGL model: it will be necessary to investigate with the
help of hybrid-PIC simulations whether and how parametric
decay competes and interacts with those kinetic
instabilities. However, depending upon plasma parameters,
saturation amplitudes of the ion-cyclotron and firehose unstable modes may be smaller than those of
Alfv\'enic fluctuations, and we argue that in those cases
the parametric decay instability should be unaffected.

\begin{figure}[htb]
\includegraphics[width=0.4\textwidth]{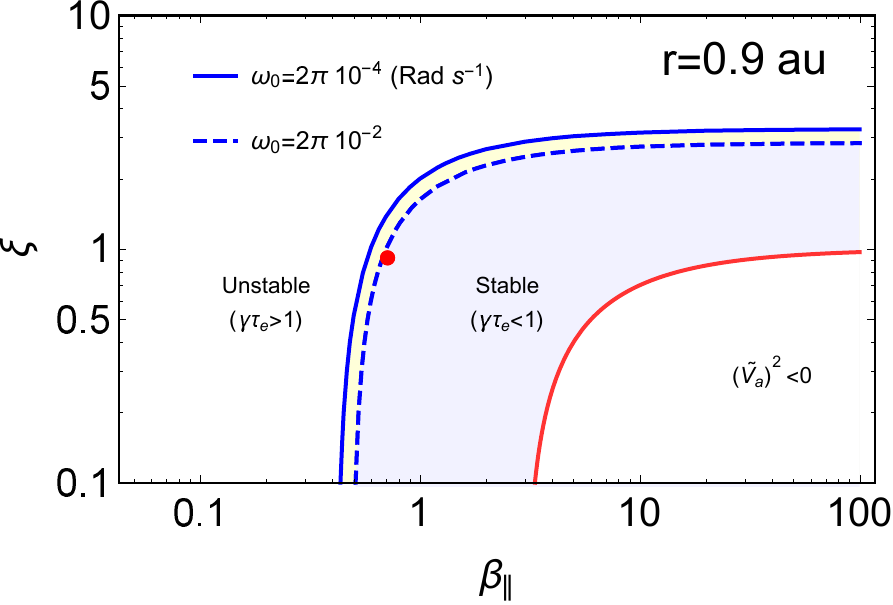}
\includegraphics[width=0.4\textwidth]{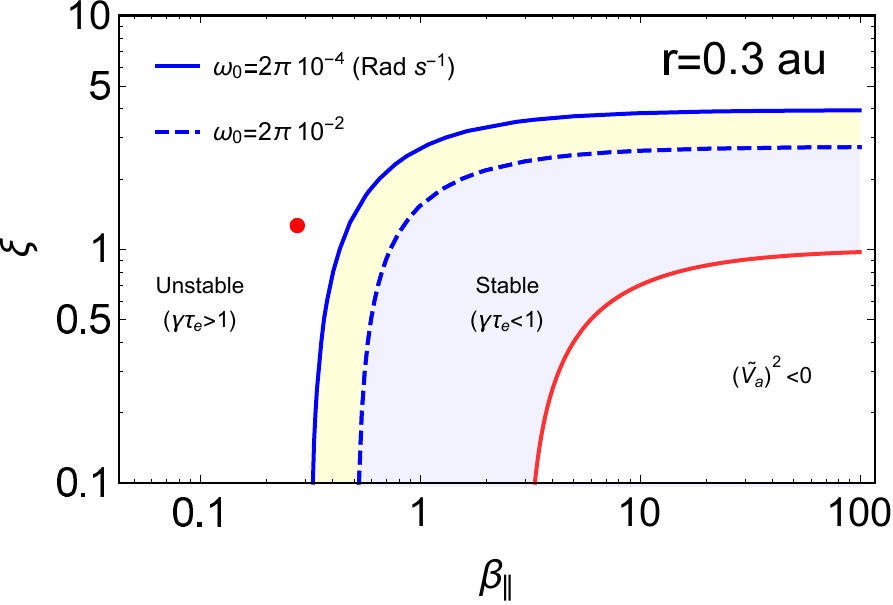}
\caption{Contours corresponding to $\gamma_m\tau_e=1$ at two heliocentric distances, $r=0.9$~AU (upper panel) and $r=0.3$~AU (lower panel), for high frequency ($\omega_{0}^{\text{\sc h}}=2\pi\times 10^{-2}$, dashed line) and low frequency ($\omega_{0}^{\text{\sc l}}=2\pi\times 10^{-4}$, solid line)  mother waves. At $r=0.9$~AU the mother wave square amplitude is $(\hat B_\bot^{\text{\sc h}})^2=0.01$ and $(\hat B_\bot^{\text{\sc l}})^2=0.1$; at $r=0.3$~AU  $(\hat B_\bot^{\text{\sc h}})^2=0.05$, and $(\hat B_\bot^{\text{\sc l}})^2=0.06$. Only parameters above the red line have been considered for parametric decay: the light blue region is unstable to parametric decay, i.e. $\gamma_m\tau_e>1$ for $\omega_0<\omega^{\text{\sc h}}$ while the white region is stable to parametric decay, i.e. $\gamma_m\tau_e<1$ for $\omega_0>\omega^{\text{\sc l}}$. The yellow region is the intermediate region where frequencies $\omega^{\text{\sc l}}<\omega_0<\omega^{\text{\sc h}}$ are unstable. The red dot indicates the location  of the average fast solar wind.}
\label{contours_SW}
\end{figure}

\acknowledgements This work was supported by grant NNX15AF34G as well as the NASA Solar Probe Plus Observatory Scientist grant. P. H. acknowledges grant 15-10057S of the Czech Foundation.

\end{document}